\documentclass{aa}  
\usepackage{natbib}
\usepackage{graphicx}
\usepackage{txfonts}
\usepackage{hyperref}
\usepackage[flushleft]{threeparttable} 
\begin{document} 

   \title{CRIRES-POP: a library of high resolution spectra in the near-infrared. III. Line identification in the K-giant 10 Leo
   }
  \author{M. Zendel\inst{1}
     \and
          T. Lebzelter\inst{1}
     \and 
          C. P. Nicholls\inst{2}
          }
  \institute{Department of Astrophysics, University of Vienna,
              T\"urkenschanzstrasse 17, A-1180 Vienna\\
              \email{thomas.lebzelter@univie.ac.at}
       \and 
       Brisbane, Australia
             }

  \date{Received 5 April 2023 / Accepted 23 August 2023 }

  \abstract
   {High-resolution spectra in the near-infrared (NIR) are an important tool for the detailed study of stellar atmospheres. The accurate identification of elements and molecules in these spectra can be used to determine chemical abundances and physical conditions in the photosphere of the observed star. Such identifications require precise line positions and strengths of both atomic and molecular features.}
   {This work focusses on the full identification of absorption lines in the NIR spectrum of the K-giant 10 Leo, including previously unidentified lines. The large number and complexity of the observed absorption lines require a deep search for potential spectral signatures to enable an unambiguous assignment to specific elements or molecular species. We aim to improve the published line lists of metals, some of which are determined by model calculations only, and many of which presently lack the completeness and accuracy of line parameters.}
   {The CRIRES-POP project provided high-resolution, high signal-to-noise ratio (S/N) spectra of several bright stars in the 1 to 5\,$\mu$m range. For the K-giant 10 Leo, a spectrum corrected for telluric absorption and with precise wavelength calibration is available. This has been analysed by comparison with model spectra and up-to-date line lists.}
   {We identified lines of 29 elements and eight molecular species. While the positions of many known lines could be confirmed, about 6\% of all lines detected in 10 Leo could not be attributed to any known feature. For CO and its isotopologues, molecular constants could be derived and several additional lines identified. We report major inconsistencies for some prominent lines. In addition, abundances for several key elements in 10 Leo are provided.}
   {}

\titlerunning{Line identification in the near infrared}
\maketitle

\section{Introduction}
Late-type stars emit the majority of their flux in the infrared, and modern infrared detectors allow for them to be studied -- within the atmospheric windows -- at a high signal-to-noise ratio (S/N) and with limited telluric absorption. 
For cool giants, a large number of molecular lines of abundant species such as CO, OH, CN, H$_2$O, and SiO populate the range between 1 and 5\,$\mu$m. 
These are of key interest as tracers of the temperature stratification of the extended atmospheres, atmospheric dynamics, nucleosynthesis, and mixing processes.
In addition, the infrared harbours a wealth of atomic lines; many of them, in particular within the atmospheric windows in the YJ band, have little blending.

However, to extract this kind of information from the stellar spectra, a complete and accurate catalogue of spectral lines that appear in that wavelength range is necessary.
Any analysis relying on synthetic spectra to derive stellar parameters and elemental abundances is limited by the quality of the line data and the knowledge of line blends. 
This is true not only for high-resolution studies, but also for work based on low- to medium-resolution spectroscopy.

High-resolution near-infrared (NIR) spectroscopy became available with the Fourier Transform Spectrometers (FTS) in the 1970s, and even more so with the recent class of NIR echelle spectrographs at 8m-class telescopes, offering access to a wide range of objects to be studied.
FTS spectra led to the first high-resolution spectral atlases with the Arcturus atlas \citep{arcturus_atlas} providing the most extensive compendium to date of line identifications of a cool star in the NIR. 
While the lasting importance of this compendium is indisputable, similar studies are needed for a wider range of stellar reference objects to constrain effects of temperature, surface gravity, and metallicity.

The CRIRES-POP project \citep{crirespop_lebzelter} obtained an observational library of high-resolution infrared stellar spectra covering the wavelength range from 1 to 5\,$\mu$m for a set of stars throughout the Hertzsprung-Russell (HR) diagram. 
The spectra have a typical S/N of 200 at a spectral resolution R=90\,000.
The first star published  with a full data reduction including a careful reanalysis of the wavelength calibration and telluric correction was the K-giant 10 Leo \citep{Nicholls_2017}. 
A summary of the stellar parameters of 10 Leo is provided in that paper. 

\begin{table}
        \centering
    \caption{Properties of 10 Leonis and Arcturus}
    \label{table:1}
        \centering
\begin{tabular}{ccccc}
\hline\hline
Property                                                &10 Leo                                         &Ref.&Arcturus&Ref.     \\
\hline
SpT                                                         &K1 IIIVAR                      &1          &K0 III       &7     \\
T$\textsubscript{eff}$          &4801$ \pm{89}$ K       &2             &4286$ \pm{30}$ K&8  \\
L                               &59.35 L$\textsubscript{$\odot$}$  &2&170 $\pm{8}$ L$\textsubscript{$\odot$}$     &9\\
R                           &14 R\textsubscript{$\odot$}&2             &25.4 $\pm{0.2}
$ R\textsubscript{$\odot$}      &8\\
M                           &2                                          &3             &1.08 $\pm{0.06}$ M\textsubscript{$\odot$}        &8\\
log \textit{g}              &2.83 $\pm{0.23}$               &2         &1.66 $\pm{0.05}$&8  \\
Distance                    &77.3 $\pm{1.5}$ pc             &4         &11.26$\pm{0.07}$pc&10\\
Age                         &4.51 $\pm{1.8}$ Gyr        &5             & $7.1_{-1.3}^{+1.5}$  Gyr                                &8\\
$[Fe/H]$                    &-0.03 $\pm{0.08}$      &2             &-0.52$\pm{0.04}$&8  \\
v\textsubscript{mic}        &1.3 km s\textsuperscript{-1}&6            &1.2 $\pm{0.11}$ km s\textsuperscript{-1}            &11\\
\hline
\end{tabular} 
\tablebib{
\citealp[(1)][]{Roman_1952}; \citealp[(2)][]{Da_Silva_2011}; \citealp[(3)][]{Mishenina_2006}; \citealp[(4)][]{Bailer_2021}; \citealp[(5)][]{Soubiran_2008}; \citealp[(6)][]{2016ApJ...823...36T}; \citealp[(7)][]{Gray_2003}; \citealp[(8)][]{Ramirez_2011}; \citealp[(9)][]{Schroder_2007};  \citealp[(10)][]{Leeuwen_2007}; \citealp[(11)][]{Kondo_2019}}
\end{table}

We briefly summarise and provide an update on the star's most important properties in the following (Table \ref{table:1}): 
10 Leo is a K1 giant with an effective temperature of 4800 K \citep{Da_Silva_2011}. 
It is a spectroscopic binary, but no lines from the companion have been found in the infrared spectrum. 
The CRIRES-POP spectrum has been corrected for the corresponding orbital motion. 
According to Gaia Early Data Release 3 (EDR3) data \citep{Bailer_2021}, the distance of 10 Leo is 77.3$\pm$1.5\,pc, which is a minor increase compared to the earlier Hipparcos distance of 75 pc used in \citet{Nicholls_2017}. 
\citet{Da_Silva_2011} found a metallicity close to solar.
However, we note that the referenced age and log g in Table \ref{table:1} for 10 Leo are not consistent with the other stellar parameters, and suggest a lower radius for 10 Leo and an age around 2.5 Gy based on the mass of 2 M$_\odot$, the reported luminosity, distance, and T\textsubscript{eff}.

10 Leo differs in some aspects from the K-giant Arcturus. Table \ref{table:1} allows for a direct comparison of the two stars. The most significant differences for our study are likely the higher temperature and the higher metallicity of 10 Leo compared to Arcturus.

In this paper, we present a detailed exploration of the spectroscopic line content in the NIR for the K-giant 10 Leo.
We present a compendium of line identifications, including a substantial number of lines not identified in the Arcturus atlas; list lines lacking identification to date; and derive parameters for the molecular bands of CO and CN between 1 and 5\,$\mu$m. 
Furthermore, we discuss the differences between the spectra of 10 Leo and Arcturus, and derive elemental abundances in 10 Leo for a set of prominent atomic species.
More extensive documentation of our data analysis is provided in \citet{Zendel_2021}.

\section{Spectral lines}

\subsection{Methods of line identification} \label{line_id}

We applied two approaches to identify the lines found in the NIR spectrum of 10 Leo. 
First, we used the Vienna atomic line database (VALD3) provided by \citet{Ryabchikova_2015} to determine all of the lines that are known or expected in the wavelength range covered. For the molecular lines of CO, CN, and OH, tables from \citet{Goorvitch_1994}, \citet{Brooke_2014}, and the High-resolution TRANsmission (HITRAN) database from \citet{Gordon_2017} were used. 
To select the lines visible for the stellar parameters of 10 Leo, we used the ‘extract stellar’ option from VALD3 which returns a list of lines with various spectral parameters such as wavenumber, log gf value, and damping constants for a particular model atmosphere. 
The chosen model atmosphere (castelli\_ap00k2\_T04750G30) for 10 Leo is based on T\textsubscript{eff}=4800 K, log g=2.83, v\textsubscript{mic}=2 km/s, and solar abundances. The minor difference to the literature value for v\textsubscript{mic} presented in Table \ref{table:1} resulted from the availability of models. 

Second, we used the Arcturus atlas by \citet{arcturus_atlas}.  
This atlas has been serving as a benchmark spectrum for high-resolution infrared spectroscopy of cool giants for decades. 
It was obtained with the FTS at Kitt Peak National Observatory and has a spectral resolution around 100\,000.
The line list for Arcturus comprises a total of 6658 absorption lines from 22 elements or ions and seven molecular species within the 1 to 5\,$\mu$m range. 
All line identifications in the Arcturus atlas were based on laboratory spectra only.
The data files for the IR spectrum of Arcturus \citep{Hinkle_1995}\footnote{downloaded from \url{ftp://ftp.noao.edu/catalogs/arcturusatlas/ir/}} cover two observation periods in 1993 and 1994. 

Most of the bad pixels in the 10 Leo spectrum had already been removed by \citet{Nicholls_2017}, but to make sure we avoid incorrect line detections, the selection criterion was set to a minimum of three neighbouring pixels being below the continuum. 
Lines in the 10 Leo spectrum were detected by an automatic search for flux values more than 0.5 \% below the local pseudo-continuum level. 
The corresponding vacuum wavelengths of the detected minima were then cross-correlated with the line lists from Arcturus, VALD3, and HITRAN.

The CRIRES-POP 10 Leo spectrum is divided into the five sections  according to the atmospheric windows. We follow this division for the presentation of findings and line lists here.
The 10 Leo spectrum comes with a number of gaps in wavelength that have no flux information, for example, the YJ bandpass has 226 gaps with sizes ranging from 0.04 nm to 2.1 nm, mostly due to strong telluric features. Details are described in \citet{Nicholls_2017}. 
We emphasise that our compendium does not include information from these gaps.

\subsection{Identified atomic lines}
Our 10 Leo NIR spectrum features a complex mixture of lines with various line shapes, a wide range of line intensities, and some discontinuities. 
The removal of telluric lines with the Molecfit software \citep{Kausch_2015, Smette_2015}  and  the applied corrections of various instrumental impacts \citep{Nicholls_2017} created some artefacts in the spectrum, mainly in the form of discontinuities of the continuum. 
This necessitated a visual inspection of the spectrum to properly identify the spectral lines.

Table \ref{table:2} provides a summary of the number of lines identified within the 10 Leo NIR spectrum sorted by bandpass. 
We note that these numbers include all components provided by the above described query of the VALD3 database. 
A significant fraction of these lines are contributions to blends.
Our approach was that whenever the spectral feature showed signs of blending, we added all lines expected within the width of the blend to our line list.
As is subsequently described in more detail in Sect. \ref{Compendium}, our line compendium takes care of this issue by adding information on the complexity of the line profile to each spectral feature.

Table \ref{table:3} lists the number of atomic lines identified, sorted by element and bandpass. 
Equivalent widths (EWs) of the strongest and weakest unblended lines are given as well. 
Values in parentheses refer to blended lines and should be regarded as upper limits of detection.
The observed elements are grouped into alpha process elements (C, O, N, Ne, Mg, Si, Ar, and Ca), odd-Z elements (Na, Al, P, K, and Sc),  iron peak elements (Ti, V, Cr, Mn, Fe, Co, and Ni),  and s-process elements (Cu, Zn, Rb, Sr, Y, Zr, Ce, and Yb).
In addition, lines from H and He have been identified.
This distinction is ambiguous in some cases, for example, Ti, Co, and Fe can be formed by the alpha process as well. 
Ca, Fe, Ti, Mg, and Y are also present in ionised form, and all others are neutral except for Sr, Ce, Dy, and Yb, which are found in ionised form only.

\begin{table*}
\caption{Overview of observed lines}              
\label{table:2}      
\centering                                     
\begin{tabular}{c c c c}         
\hline\hline                       
wavelength& listed lines & \multicolumn{2}{c}{observed lines\textsuperscript{b}} \\  
range     & VALD3\textsuperscript{a}  & identified  &  unidentified     \\
\hline                                   
YJ      &       4321    &       3664    &       433     \\
H       &       6586    &       4438    &       303     \\
K       &       4166    &       3018    &       196     \\
L       &       1900    &       1304    &       415     \\
M       &       3080    &       2577    &       84      \\
\hline
SUM:&   20053   &       15001   &       1431\\
\hline                                             
\end{tabular}
\begin{tablenotes}
 \item \textsuperscript{a} From VALD3 stellar extraction corrected for gaps
\item \textsuperscript{b} Includes all observed lines of the 10 Leo NIR spectrum
 \end{tablenotes}
\end{table*}
\begin{table*}
\caption{10 Leo: Identified lines per element}              
\label{table:3}      
\centering                                      
\begin{tabular}{l r r r r r r r r r r}         
\hline\hline                        
Element     & YJ &  H & K  & L  & M & total & strongest & EW     & weakest   & EW    \\
            &    &    &    &    &   &       & line [nm] & [m\AA] & line [nm] & [m\AA]\\
\hline
\ion{H}{I}  & 3  & 6  & 1  & 2  & 1 & 13    & 1681.11   & 232    & 1556.07   & (6)   \\
\ion{He}{I} & 1  & 0  & 0  & 0  & 0 & 1     & 1083,33   & 70     &           &       \\
\hline
\multicolumn{11}{c}{Alpha-process}\\
\hline
\ion{C}{I}  & 40 & 83 & 19 & 43 & 0 & 185   & 1069.42   & 117    & 3777.32   & 2     \\
\ion{Mg}{I} & 41 & 85 & 76 & 75 & 10& 287   & 3677.19   & 271    & 1030.21   & 12    \\
\ion{Mg}{II}& 2  &  0 &  0 &  0 &  0&   2   & 1095.48   & 11     & 1091.84   & (3)   \\
\ion{Si}{I} & 135& 126& 140& 132& 16& 549   & 2206.87   & 305    & 1080.01   & 2     \\
\ion{S}{I}  & 11 & 21 & 19 & 2  & 0 &  53   & 1046.22   & 58     & 2289.35   & 8     \\
\ion{Ca}{I} & 54 & 33 & 48 & 21 & 6 & 162   & 2263.10   & 278    & 2194.87   & 3     \\
\ion{Ca}{II}& 5  & 2  & 4  & 0  & 0 &  11   & 1184.22   & 114    & 985.75    & 7     \\
\hline
\multicolumn{11}{c}{odd-Z}\\
\hline
\ion{Na}{I} & 21 & 7   & 17 & 16 & 3 &  64   & 2208.97  & 331    & 991.87    & (3)   \\ 
\ion{Al}{I} & 10 & 10  & 7  & 9  & 1 &  37   & 1312.70  & 479    & 3601.13   & 6     \\
\ion{P}{I}  & 8  & 8   & 0  & 0  & 0 &  16   & 1058.45  & 16     & 1068.43   & 6     \\
\ion{K}{I}  & 7  & 3   & 2  & 7  & 0 &  19   & 1177.61  & 158    & 2194.87   & 3    \\
\ion{Sc}{I} & 0  & 0   & 9  & 0  & 0 &  9    & 2207.13  & 34     & 2173.64   & 2     \\
\hline
\multicolumn{11}{c}{Iron-peak}\\
\hline
\ion{Ti}{I} &106 &74  & 29  & 9  & 3 & 221   & 964.09   & 192    & 998.40    & 2     \\
\ion{Ti}{II}&3   &7   & 0   & 0  & 0 & 10    & 1587.82  & 68     & 969.39    & (8)   \\
\ion{V}{I}  &10  &16  & 8   & 0  & 0 & 34    & 1290.47  & 21     & 1318.86   & 2     \\  
\ion{Cr}{I} &51  &20  & 10  &10  & 2 & 93    & 1101.85  & 77     & 1295.06   & 4    \\
\ion{Mn}{I} &13  &36  & 6   & 6  & 0 & 61    & 1332.26  & 322    & 3473.88   & 6    \\ 
\ion{Fe}{I} &440 &1205&372  &393 & 9 &2419   & 1188.61  & 325    & 1217.44   & 2     \\
\ion{Fe}{II}&4   &5   & 1   & 0  & 0 & 10    & 1086.56  & 13     & 1112.85   & (3)   \\
\ion{Co}{I} &14  &26  & 4   & 0  & 0 & 44    & 1676.22  & 82     & 2207.29   & 3    \\
\ion{Ni}{I} &43  &93  &31   & 41 & 1 &209    & 1631.50  & 144    & 1240.51   & 3    \\
\hline
\multicolumn{11}{c}{s-process}\\
\hline
\ion{Cu}{I} &0   &4   & 0   & 0  & 0 &  4    & 1601.00   & (27)  & 1601.09   & (4)   \\    
\ion{Zn}{I} &3   &0   & 0   & 0  & 0 &  3    & 1305.71   & 17    & 1105.73   & (9)   \\
\ion{Rb}{I} &0   &2   & 0   & 0  & 0 &  2    & 1529.37   & (6)   &           &       \\
\ion{Sr}{II}&3   &0   & 0   & 0  & 0 &  3    & 1033.01   & 232   & 1003.94   & 119   \\
\ion{Y}{I}  &0   &1   & 5   & 0  & 0 &  6    & 1766.80   & 17    & 2255.00   & 5     \\
\ion{Y}{II} &5   &0   & 0   & 0  & 0 &  5    & 1024.80   & 11    & 1018.93   & 6     \\
\ion{Zr}{I} &3   &0   & 0   & 0  & 0 &  3    & 982.53    & 3     & 1166.11   & 2     \\
\ion{Ce}{II}&36  &15  & 3   & 0  & 0 &  54   & 1708.20   & 56    & 1258.57   & 5     \\
\ion{Dy}{II}&1   &0   & 0   & 0  & 0 &  1    & 1052.63   & 7     &           &       \\
\ion{Er}{II}&1   &0   & 0   & 0  & 0 &  1    & 1106.26   & (11)  &           &       \\
\ion{Yb}{II}&0   &1   & 0   & 0  & 0 &  1    & 1650.28   & (10)  &                   &        \\
\hline
\end{tabular}
\begin{tablenotes}
 \item Values in parentheses refer to blended lines and should be regarded as upper limits of detection.
 \end{tablenotes}
\end{table*}

\subsection{Molecular lines and bandheads}
Many molecular species form band structures in the infrared with distinct patterns for the line positions and strengths.
While this in principle eases the identification of molecular lines, some bands are blended with others complicating the analysis.
The abundant molecular lines in the 10 Leo spectrum mainly stem from CO, CN, and OH molecules. 
Lines from the isotopologues \element[][13]{C}\element[][16]{O}, \textsuperscript{12}C\element[][17]{O}, and \textsuperscript{12}C\element[][18]{O} are easily detectable throughout the spectrum.
However, no lines of C\element[][15]{N} could be detected based on a search using the catalogue of CN lines from \cite{Brooke_2014}. 
Each band was fitted with a fourth order polynomial to locate the individual transition lines and to ensure consistency in transition assignment. 
Fundamental bands ($\Delta\nu$=1) of CO are observed in the M band, first overtone bands ($\Delta\nu$=2) in the K band, and second overtone bands ($\Delta\nu$=3) in the H band. 
A total of 105 CO bands (64 in M, 23 in K, and 18 in H) were evaluated which comprises about 4500 CO lines (almost twice the number of CO lines identified in the Arcturus atlas).
Table \ref{table:4} summarises all fundamental bands from which lines could be identified in the 10 Leo spectrum. 
The wavelengths given for the R0 and P1 lines of each band were computed from the polynomials deduced for each band.
Central depths (cdepth) and EWs are given for the strongest detected line in each band.
\begin{table*}
    \centering
        \caption{Fundamental CO bands}
        \label{table:4}
\begin{threeparttable}  
\begin{tabular}{llrcclrcc}
\hline\hline
        &\multicolumn{4}{c}{R-branch}&\multicolumn{4}{c}{P-branch}\\
        \cline{2-9}
&&&&&&&&\\
Band-ID                     &$\lambda$ (R0)$\tnote{\textsuperscript{a}}$&Transition\tnote{\textsuperscript{b}}   & (1-F)    &   cdepth\tnote{\textsuperscript{c}}    &$\lambda$ (P1)$\tnote{\textsuperscript{a}}$ &Transition \tnote{\textsuperscript{b}}&(1-F)& cdepth\tnote{\textsuperscript{c}} \\
&nm &&&&nm &&&\\
\hline
&&&&&&&&\\
CO 1-0  &       4657.49\tnote{d}        &       R6      &       0.587   &       0.482   &       4674.15\tnote{d}         &       P11     &       0.517   &       0.487    \\
CO 2-1  &       4715.72         &       R14     &       0.506   &        0.492   &       4732.65         &       P23     &       0.513   &       0.477    \\
CO 3-2  &       4775.28         &       R22     &       0.472   &       0.474   &       4792.48         &       P30     &       0.424   &       0.443    \\
CO 4-3  &       4836.21         &       R29     &       0.419   &       0.441   &       4853.69         &       P17     &       0.397   &       0.416    \\
CO 5-4  &       4898.54\tnote{d}        &       R34     &       0.373   &       0.400   &       4916.30\tnote{d}         &       P7      &       0.336   &       0.351    \\
CO 6-5  &       4962.33         &       R31     &       0.332   &       0.365   &       4980.38\tnote{d}        &       P20     &       0.317   &       0.342    \\
CO 7-6  &       5027.62\tnote{d}        &       R13     &       0.264   &       0.318   &       5045.98         &       P4      &       0.232   &       0.262    \\
CO 8-7  &       5094.45         &       R25     &       0.253   &       0.301   &       5113.12\tnote{d}         &       P8      &       0.248   &       0.256    \\
CO 9-8  &       5162.90\tnote{d}        &       R24     &       0.219   &       0.271   &               &               &               &                 \\
CO 10-9         &       5232.99\tnote{e}        &       R19     &       0.186   &       0.235   &                &               &               &                \\
CO 11-10        &       5304.80\tnote{e}        &       R27     &       0.168   &       0.210   &                &               &               &                \\
CO 12-11        &       5378.53\tnote{e}        &       R45     &       0.121   &       0.151   &                &               &               &                \\
CO 13-12        &       5454.06\tnote{e}        &       R35     &       0.122   &       0.086   &                &               &               &                \\
CO 14-13        &       5531.54\tnote{e}        &       R55     &       0.043   &       0.051   &                &               &               &                \\
\element[][13]{C}O 1-0  &       4762.56         &       R18     &       0.367   &        0.285   &       4779.22         &       P25     &       0.307   &       0.266          \\
\element[][13]{C}O 2-1  &       4820.76\tnote{d}        &       R30     &       0.344   &       0.280   &       4837.67         &       P18     &       0.317   &       0.248    \\
\element[][13]{C}O 3-2  &       4880.25\tnote{d}        &       R36     &       0.265   &       0.248   &       4897.42         &       P29     &       0.247   &       0.221    \\
\element[][13]{C}O 4-3  &       4941.07         &       R37     &       0.248   &       0.212   &       4958.52\tnote{d}         &       P25     &       0.231   &       0.189    \\
\element[][13]{C}O 5-4  &       5003.26\tnote{d}        &       R10     &       0.177   &       0.130   &       5020.99\tnote{d}         &       P17     &       0.235   &       0.139    \\
\element[][13]{C}O 6-5  &       5066.87         &       R51     &       0.141   &       0.085   &       5084.89\tnote{d}         &       P5      &       0.103   &       0.040    \\
\element[][13]{C}O 7-6  &       5131.95 &       R33     &       0.145   &       0.078   &       &               &                &                \\
\element[][13]{C}O 8-7  &       5198.49\tnote{d}        &       R32     &       0.109   &       0.048   &                &               &               &                \\
\element[][13]{C}O 9-8  &       5266.64\tnote{e}        &       R26     &       0.079   &       0.029   &                &               &               &                \\
\element[][13]{C}O 10-9         &       5336.37\tnote{e}&       R36     &       0.064   &       0.016   &                &               &               &                \\
\element[][13]{C}O 11-10        &       5407.79\tnote{e}&       R28     &0.053         &0.013      &           &               &               &                 \\
C\element[][18]{O} 1-0  &       4771.56         &       R22     &       0.108   &        0.158   &       4788.22         &       P17     &       0.096   &       0.125    \\
C\element[][18]{O} 2-1  &       4829.92         &       R19     &       0.103   &        0.140   &       4846.67 &       P17     &       0.049   &       0.114    \\
C\element[][18]{O} 3-2  &       4889.23 &       R29     &       0.071   &       0.114   &       4906.41 &       P15     &       0.069   &       0.079    \\
C\element[][18]{O} 4-3  &       4950.04 &       R45     &       0.035   &       0.062   &       4967.50 &       P26     &       0.046   &       0.062    \\
C\element[][18]{O} 5-4  &       5012.22 &       R6      &       0.024   &       0.020   &       5029.94\tnote{d}         &       P6      &       0.167   &       0.016    \\
C\element[][18]{O} 6-5  &       5075.83         &       R4      &       0.017   &       0.008   &                &               &               &                \\
C\element[][18]{O} 7-6  &       5140.88\tnote{d}        &       R29     &       0.037   &       0.018   &                &               &               &                \\
C\element[][18]{O} 8-7  &       5207.42 &       R17     &       0.011   &       0.008   &                &               &               &                \\
C\element[][17]{O} 1-0  &       4716.96 &       R6      &       0.086   &        0.017   &       4733.62 &       P21     &       0.111   &        0.033  \\
C\element[][17]{O} 2-1  &       4775.17 &       R11     &       0.103   &       0.025   &       4792.09 &       P23     &       0.088   &       0.030    \\
C\element[][17]{O} 3-2  &       4834.69 &       R22     &       0.070   &       0.026   &       4851.88 &       P20     &       0.078   &       0.020    \\
C\element[][17]{O} 4-3  &       4895.57 &       R30     &       0.061   &       0.017   &       4913.02 &       P14     &       0.036   &       0.011    \\
C\element[][17]{O} 5-4  &       4957.81\tnote{d}        &       R29     &       0.031   &       0.010   &                &               &               &                \\
C\element[][17]{O} 6-5  &       5021.50\tnote{d}        &       R15     &       0.020   &       0.006   &                &               &               &                \\
\hline        
\end{tabular}
\begin{tablenotes}
  \item \textsuperscript{a}Fit value;
\textsuperscript{b}strongest, least blended line selected;
\textsuperscript{c}value from VALD stellar extraction;
\textsuperscript{d}line absent due to a void in the CRIRESPOP spectrum; 
\textsuperscript{e}line outside the wavelength range.
\end{tablenotes}
\end{threeparttable} 
\end{table*}

Table \ref{table:5} lists all of the identified band heads for the CO molecule in the studied spectral range, including molecular band data. 
The band heads occur in the R branches as a result of the increase in the distortion term with higher J values. 
Their intensity decreases steeply from the fundamentals to the second overtones. 
However, CO bandheads for the fundamentals  appear at higher J values (J$\approx$88) than for the first overtones (J$\approx$51) and second overtones (J$\approx$34), which enhances the possibility of detecting overtone bands again.
We note that we could not detect any bandheads for the two isotopologues C$^{17}$O and C$^{18}$O in the spectrum of 10 Leo.

\begin{table*}
    \centering
        \caption{Identified CO bandheads}
        \label{table:5}
\begin{tabular}{lcccccc}
\hline\hline
&\multicolumn{3}{c}{\element[][12]{C}O}&\multicolumn{3}{c}{\element[][13]{C}O}\\
\cline{2-7}     
Transition&     $\lambda$ (nm)& Literature\tnote(\textsuperscript{a}&difference (nm)&$\lambda$(nm)&     Literature\tnote(\textsuperscript{b}&difference (nm)    \\
\hline                                                                                                  
fundamental     &               &                   &           &               &                &               \\
4-3         &outside    &4465.21    &           &       4561.66 &4561,67    &-0.01               \\
5-4         &outside    &4524.63    &           &       4620.95 &4620,95         &-0.01          \\
6-5         &   4585.45 &4585.45    &0.00       &       4681.60 &4681,60    &0.00                \\
7-6         &   4647.71 &4647.72    &-0.01      &       4743.64 &4743,66    &-0.02               \\
8-7         &   4711.47 &4711.47    &0.00       &       4807.13 &4807,16         &-0.03          \\
9-8         &   4776.76 &4776.75    &0.01       &       4872.11 &4872,14         &-0.03          \\
10-9    &       gap     &4843.63    &           &       4938.64 &4938,68         &-0.04          \\
11-10   &       4912.14 &4912.14    &0.00       &                   &               &            \\
12-11   &       4982.35 &4982.36    &-0.01      &               &                   &            \\
13-12   &       5054.33 &5054.33    &0.00       &                   &               &            \\
\hline                                                                                                  
1st overtone&           &                   &       &               &               &            \\
2-0     &       2293.52 &       2293.52 & 0.00  &       2344.85 &       2344.80 &       0.05    \\
3-1         &   2322.65 &       2322.66 &-0.01  &       2373.94 &       2373.90 &       0.04    \\
4-2     &       2352.46 &       2352.46 & 0.00  &       2403.69 &       2403.70 &       -0.01   \\
5-3     &       2382.94 &       2382.95 &-0.01  &       2434.12 &       2434.10 &       0.02    \\
6-4         &   2414.14 &       2414.14 & 0.00  &       2465.24 &       2465.20 &       0.04    \\
7-5     &       2446.06 &       2446.08 &-0.02  &       2497.08 &       2497.10 &       -0.02    \\
8-6     &       2478.76 &       2478.76 & 0.00  &               &                &               \\
9-7         &   2512.22 &       2512.23 &-0.01  &               &                &               \\
10-8    &               &               &               &                   &            &                   \\
\hline                                                                                                  
2nd overtone&       &           &               &               &                &                   \\
3-0         &   1558.16 &       1558.16 &0.00   &                   &            &                   \\
4-1     &       1577.95 &       1577.95 &0.00   &               &                   &                \\
5-2         &   1598.18 &       1598.19 &-0.01  &                   &            &                   \\
6-3         &   1618.89 &       1618.89 &0.00   &                   &            &                   \\
7-4         &   1640.08 &       1640.08 &0.00   &                   &            &                   \\
8-5         &   1661.77 &       1661.77 &0.00   &                   &            &                   \\
9-6         &   1683.97 &       1683.97 &0.00   &                   &            &                   \\
10-7    &       1706.71 &       1706.71 &0.00   &                   &            &                   \\
11-8    &       1729.99 &       1729.99 &0.00   &                   &            &                   \\
12-9    &       1753.84 &       1753.84 &0.00   &                   &            &                   \\
13-10   &       1778.28 &       1778.28 &0.00   &                   &            &                   \\
\hline        
\end{tabular}
\begin{tablenotes}
  \item[a] \textsuperscript{a}\citet{Goorvitch_1994}
  \item[b] \textsuperscript{b}\citet{13CO} 
\end{tablenotes}
\end{table*}
Fundamental bands from electronic transitions for CN are located in the YJ and H regions, whereas the first overtone bands from the electronic ground state are found in the K band. 
For CN, 138 bands (61 in YJ, 41 in H, and 36 in K) have been evaluated resulting in the assignment of about 5158 individual CN lines matched with the VALD3 {\tt stellar extraction} option (see \ref{line_id}) and the Arcturus atlas.
The maximum central depth for all observed molecular lines within a band was recorded at J $\approx$31 with the exception of the satellite bands of CN, which have the maximum at J $\approx$10. 
For the OH molecule, 32 bands have been identified and evaluated (16 in H and 16 in L).

Weak signatures of CH and SiO have been found. 
The NH molecule had been detected in the Arcturus spectrum in a region from 2891.00 nm to 3397.94 nm. 
This region is not covered by the 10 Leo spectrum except for a small part starting at 3367.02 nm. 
Within this region two lines at 3373.06 nm and 3397.94 nm could be attributed to NH.
In the L band, we also found seven lines that could be assigned to the HCl molecule.
For HF, two lines (1-0 R3 and 1-0 R9) could be detected.

\subsection{Unidentified lines}\label{Unidentified lines}

After carefully attributing known atomic and molecular transitions to the detected spectral features in 10 Leo, about 1400 lines or, in the case of line blends, line components were left unidentified. 
We mark a line or line component as unidentified if we could not find any candidate transition within 0.1 nm of the observed wavelength position that was not already attributed to a line or line component in the 10 Leo spectrum.
The fraction of unidentified lines is $\approx$ 9\% of all observed lines at the 0.5$\%$ sensitivity level in the 10 Leo spectrum. Throughout the spectrum, we found more than 350 unidentified lines that are isolated lines with no obvious traces of blending components.
Most of the unidentified lines, however, are components within line blends.

Table \ref{table:6} shows the number of such lines for each band  sorted in intervals of EWs of $<$ 10 mÅ, $\geq$10 mÅ and $<$ 50 mÅ, $\geq$50 mÅ and $<$ 100 mÅ, and $\geq$100 mÅ, respectively.
The numbers in the last category clearly show that there are several strong lines in the NIR range, which have not been correctly identified yet.
The Apache Point Observatory Galactic Evolution Experiment (APOGEE) project has also published a list of unidentified lines \citep{2021Apogee}, which is limited to the range between 1547 and 1693 nm.
We find 33 unidentified lines in common with their list.

Very recently, \citet{Peterson_2022} published a list of new \ion{Fe}{I} line identifications which included the infrared range.
We cross-matched their line positions with our list of unidentified lines and found 162 lines to agree within $\pm$0.05 nm.
Since many of these lines are weak with EWs <10 nm in the spectrum of 10 Leo, it is difficult to determine their correctness without including them in spectral synthesis computations, which is beyond the scope of this paper.
For the quite strong line at 4574.46 nm (EW = 644 mÅ), we suspect an accidental agreement, because other lines in the list of \citet{Peterson_2022} with similar line parameters do not show a strong line in 10 Leo and because these authors note that the new lines identified in their paper are typically weak.
However, a tentative identification of most lines in common seems reasonable.

\begin{table*}
    \centering
        \caption{Number of unidentified lines}
        \label{table:6}
\begin{threeparttable}          
\begin{tabular}{lrrrrrr}
\hline\hline
&\multicolumn{6}{c}{Band}\\
\cline{2-7}     
        EW-Interval                     & YJ  &  H  & K  &  L   &  M & All\\
\hline  
$<$10 mÅ                                   & 248 & 106 &  96 & 135 &  1 & 586\\
$\geq$10 mÅ $<$50 mÅ      & 165 & 129 &  79 & 229     & 32 & 634\\    
$\geq$50 mÅ $<$100 mÅ       &  20 &  35 &  12 &        40     & 23 & 130\\    
$\geq$ 100mÅ                           &       0 &  33 &   9 &  11 & 28 &  81\\ 
\hline\hline
Total                       & 433 & 303 & 196 &  415 & 84 &1431\\
\hline
\end{tabular}
\end{threeparttable}    
\end{table*}

Table \ref{table:7} lists the most intense unidentified lines (EW$>$50 m{\AA}) for the YJ band along with possible blending transitions. 
The rightmost column gives the reference suggesting the blending components.
Similar lists for the other bands with an EW>100 m{\AA} are provided in the Appendix (Tables \ref{table:11} and \ref{table:12}).
For the YJ band, there are no unidentified lines or line components with an EW exceeding 100 m\AA.

\begin{table*}
\caption{Most intense unidentified lines with an EW>50 mÅ in the YJ band and possible contributors}
\label{table:7}
\begin{tabular}{l c l c}
\hline\hline
wavelength&EW   &possible blends                                &reference\\
(nm)      &mÅ   &                                               &         \\
\hline
1113.69 &       79      &       \ion{Co}{I} 1113.68, \ion{Fe}{I} 1113.69,
\ion{V}{i} 1113.70, \ion{Ca}{I} 1113.70                         &       (1)     \\
1120.30 &       64      &                                                   &            \\
1121.40 &       55      &       \ion{Fe}{I} 1121.39, \ion{Fe}{II} 1121.41       &       (1)     \\
1127.28 &       67      &       \ion{Fe}{II} 1127.28, \ion{N}{I} 1127.28?       &       (1)     \\
1131.31 &   89  &   \ion{Cr}{I} 1131.38                         &  (1),(2)\\
1131.51 &   97  &   (CN 0-0Q(1)38.5 at 1131,47 subtracted)      &       \\
1132.25 &       58      &       doublet \ion{Fe}{II} 1132.24, 1132.25           &       (1)     \\
1132.40 &       66      &       \ion{Fe}{II} 1132.39, CN 1-1R(2)25.5            &       (1)     \\
1139.26 &       56      &       \ion{Co}{I} 1139.27, \ion{Cr}{II} 1139.28, 
(CN 1-1Q(2)21.5 at 1139.27 subtracted)          &       (1)     \\
1141.54 &       81      &       \ion{N}{I} 1141.54?, \ion{Cr}{II} 1141.52       &       (1)     \\
1144.07 &       54      &       \ion{Ti}{I} 1144.05, \ion{Mn}{II} 1144.04       &       (1)     \\
1146.27 &       69      &       \ion{N}{I} 1146.27?, (CN 1-1R(1)37.5 at 1146.27 subtracted)     &               \\
1150.19 &       78      &                                                                                           &            \\
1165.18 &       61      &       \ion{Fe}{II} 1165.18, \ion{Al}{I} 1165.22,
\ion{Mn}{ii} 1165.24, (CN 1-1P(2)30.5 at 1165.20 subtracted)    &       (1)     \\
1198.77 &   60  &   strong shoulder of \ion{Si}{I} 1198.75      &       \\
1261.72 &       53      &       \ion{Fe}{II} 1261.69, \ion{C}{I} 1261.75        &       (1)     \\
1263.17 &       53      &       \ion{Ti}{I}  1263.17, \ion{Cr}{II} 1263.18      &       (1)     \\
1308.06 &       86      &       \ion{Fe}{II} 1308.09, \ion{Fe}{I} 1308.09, 
\ion{Ti}{i} 1308.08 (cdepth 0.253) subtracted   &       (1)(2)  \\
\hline
\end{tabular}
\tablebib{(1)\citet{Ryabchikova_2015},(2)\citet{Hinkle_1995}}
\end{table*}

\subsection{Incorrect transitions in VALD3}
A total of 5718 lines predicted by the spectral extract mode from VALD3 (see \ref{line_id} for the stellar parameters of 10 Leo) could not be observed in the CRIRES-POP spectrum. 
The most prominent ones (predicted central depth exceeding 10\,\% of the continuum level) missing in the observed spectrum are listed in Table \ref{table:8} for the YJ band.
This list includes lines from elements of high astrophysical importance such as \ion{Mg}{I} and \ion{Fe}{I}.
Assuming wrong line parameters in these cases, these transitions may correspond to some of the unidentified lines mentioned above. 
Similar tables for the remaining bands can be found in the Appendix, Tables \ref{table:13}, \ref{table:14}, and \ref{table:15}.

\begin{table}
\caption{Lines with cdepth>0.10 not present in observed YJ band}             
\label{table:8}      
\centering                             
\begin{threeparttable}  
\begin{tabular}{c c c c}      
\hline\hline                        
Wavelength&&Absorptance\textsuperscript{a}&Synthetic\textsuperscript{b}\\ 
(nm)&species&1-F&cdepth \\
\hline
964.20  &       \ion{Fe}{I}     &       0.016   &       0.176   \\
969.90  &       \ion{Ca}{I}     &       -0.022  &       0.148   \\
974.16  &       \ion{Fe}{I}     &       0.018   &       0.107   \\
977.10  &       \ion{Fe}{I}\textsuperscript{c}  &       0.136   &       0.113   \\
982.43  &       \ion{Ca}{I}     &       -0.009  &       0.213   \\
983.08  &       \ion{Ca}{I}     &       0.005   &       0.102   \\
1038.78 &       \ion{Si}{I}     &       0.007   &       0.102   \\
1062.08 &       \ion{Fe}{I}     &       0.012   &       0.107   \\
1083.49 &       \ion{Ca}{I}     &       -0.005  &       0.152   \\
1096.03 &       \ion{Mg}{I}\textsuperscript{d}  &       0.415   &       0.323   \\
1096.03 &       \ion{Mg}{I}\textsuperscript{d}  &       0.321   &       0.416   \\
1098.75 &       \ion{Si}{I}\textsuperscript{e}  &       0.316   &       0.281   \\
1166.59 &       \ion{Fe}{I}     &       -0.015  &       0.141   \\
1181.48 &       \ion{Ca}{I}     &       0.017   &       0.325   \\
1182.42 &       \ion{Ca}{I}     &       0.002   &       0.212   \\
1187.74 &       \ion{La}{II}&   0.007   &       0.110   \\
1274.37 &       \ion{Ca}{I}     &       -0.003  &       0.160   \\
1312.07 &       \ion{Ca}{I}     &       0       &       0.106   \\
\hline
\end{tabular}
\begin{tablenotes}
 \item\textsuperscript{a}observed (1-F) at wavelength $\lambda\textsubscript{0}$ 
 \item\textsuperscript{b}VALD {\tt stellar extraction} mode (T\textsubscript{eff}=4800 K, log g= 2.67, v\textsubscript{mic}=2 km\textsuperscript{-1}, scaled to solar values).
 
 \item\textsuperscript{c}The observed absorptance can be explained by \ion{Ti}{I} and \ion{Si}{I}.
\item\textsuperscript{d}Two Mg lines predicted. The observed absorptance can be explained by CN only. 
\item\textsuperscript{e}The observed absorptance can be explained by CN only.
\end{tablenotes}
\end{threeparttable}    
\end{table}

\subsection{Comparison of the 10 Leo and Arcturus spectrum}
A first comparison of the high-resolution spectra of the two stars was presented in \citet{Nicholls_2017}. 
10 Leo and Arcturus are both K stars, both with luminosity and radius values that place them on the red giant branch. 
Arcturus is $\approx$ 500 K cooler than 10 Leo and has a lower metallicity. 
It is much older than 10 Leo and has only half its mass. 
Accordingly, they also differ in log \textit{g} value, radius, and luminosity. 
\citet{Blum_2003} and \citet{Schultheis_2016} have shown that the molecular bands of CO increase in strength in K and early M giants with decreasing temperature, while there is no obvious dependence on metallicity.
Model calculations by \citet{aringer_et_al_2016} suggest that OH lines in the NIR show a similar behaviour. 
\citet{Rich_Origlia_2005} found an increase in CO band strengths with decreasing log g, which is not seen in individual OH lines they investigated. The low sensitivity of the OH lines to changes in the surface gravity was confirmed by \citet{Lebzelter_2008} for the H band. Based on these findings and the differences in the stellar parameters between the two stars, we expect the Arcturus spectrum to tentatively feature more pronounced and stronger molecular lines than 10 Leo. For atomic lines, the differences in metallicity and stellar parameters between the two stars do not allow for a general prediction.

Looking at the number of identified lines, there are 5171 lines in common between the Arcturus atlas and our 10 Leo line compendium, which also have have an assigned transition from VALD3. 
Among these are 1084 (21\%) isolated lines, that is, lines with no obvious blends; a similar number of lines are part of blends, that is, showing two clearly distinguished line minima with no other obvious blending components; and the remaining lines, which are more affected by line blends.

\begin{figure}
\includegraphics[width=88mm, height=50mm]{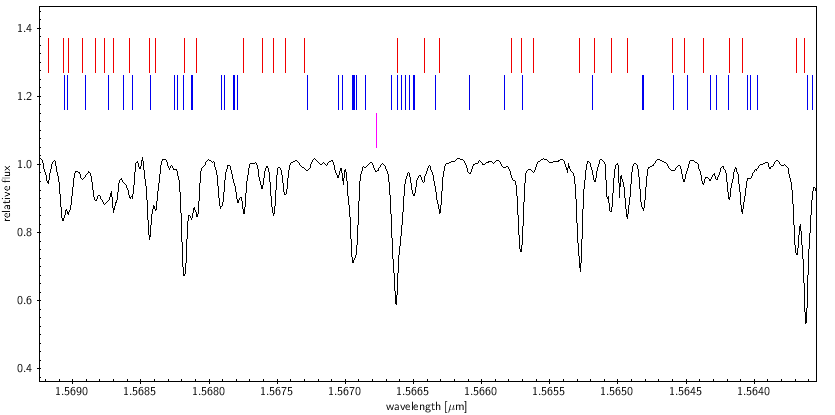}  
\caption{Example section from the 10 Leo spectrum. Identified lines are marked by vertical strokes. The top row (red) corresponds to line identifications in common with the Arcturus atlas. The second row (blue) are identifications attributed based on our model calculations. The single magenta line in the third row indicates an as of yet unidentified feature.}
\label{figure:atlas2}
\end{figure}
In addition, we identified 9183 lines 
in 10 Leo with an assigned transition from VALD3 that were not identified in the Arcturus atlas, with 9\% of them being isolated lines, meaning that most of these additional identifications are components of line blends.
Figure \ref{figure:atlas2} shows an example section with lines identified in our compendium marked.
Interestingly, there are 251 lines 
in 10 Leo with an assignment from the Arcturus atlas that do not show up in the VALD3 database.
Furthermore, 292 lines have an identification in the Arcturus atlas, but are missing from the 10 Leo line list. 
The latter differences are partly due to slight deviations in the covered wavelength range, defect or saturated spectral regions in either of the two spectra, and partly due to different line strengths excluding some lines from being selected.

We compared EWs for all lines identified in both spectra. The largest differences in central depth are noted for \ion{Ti}{I}, \ion{Sc}{I}, \ion{V}{I}, OH, and CO, which show a lower EW in 10 Leo. Some differences between the isotopologues of CO are observed. 
In 10 Leo, we see stronger lines of  $\textsuperscript{12}$C$\textsuperscript{17}$O compared to Arcturus, whereas $\textsuperscript{13}$C$\textsuperscript{16}$O bands are weaker in 10 Leo than in Arcturus.
For \ion{Fe}{I}, \ion{S}{I}, \ion{C}{I,} and CN, lines are stronger in 10 Leo. We come back to this point in the context of abundance determination in Sect.\,\ref{sect:abundances}.

\begin{figure}
\includegraphics[width=88mm, height=60mm]{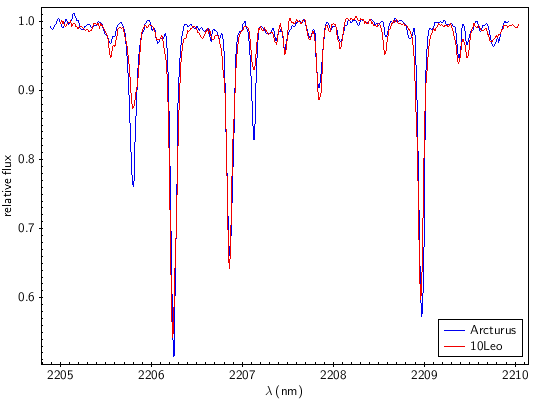}  
\caption{Overlay spectra 10 Leo (red) and Arcturus (blue). See text for details.}
\label{figure:3}
\end{figure}

Figure \ref{figure:3} shows a typical spectral section with six major lines at 2205.82 nm (\ion{Sc}{I}), 2206.24 nm (\ion{Na}{I}), 2206.87 nm (\ion{Si}{I}), 2207.13 nm (\ion{Sc}{I}), 2207.86 nm (\ion{Si}{I}), and 2208.97 nm (\ion{Na}{I}) as an overlay between 10 Leo and Arcturus. 
The most obvious difference between the two spectra are the \ion{Sc}{I} lines with a significantly enhanced line depth in Arcturus relative to 10 Leo. 
Minor differences can be seen for the \ion{Na}{I} and \ion{Si}{I} lines. 
Table \ref{table:9} shows the ratios of EWs between the two stars for various metals and molecules.

 \begin{table}
    \centering
        \caption{Ratios of EWs (Arcturus and 10 Leo)} 
        \label{table:9}
        \centering
\begin{tabular}{ccc}
\hline\hline
species       & Ratio                   & $\#$ evaluated \\
                  & Arcturus/10 Leo     &       lines        \\
\hline     
CO                &1.73  $\pm$  0.23    &       35      \\
CH                &1.37  $\pm$  0.31    &       10      \\
                                        
\ion{C}{I}        &0.76  $\pm$  0.15    &       9       \\
CN                &0.83  $\pm$  0.13    &       171     \\
                                        
OH                &3.26  $\pm$  0.43    &       37      \\
\ion{Na}{I}   &0.94  $\pm$  0.03        &       3       \\
                                        
\ion{Mg}{I}   &1.10  $\pm$  0.14        &       20      \\
\ion{Al}{I}   &1.12  $\pm$  0.02        &       10      \\
                                        
\ion{Si}{I}   &0.94  $\pm$  0.10        &       51      \\
\ion{K}{I}        &1.13  $\pm$  0.13    &       4       \\
                                        
\ion{Ca}{I}   &1.01  $\pm$  0.11        &       22      \\
\ion{Sc}{I}   &1.90  $\pm$  0.68        &       3       \\
                                        
\ion{Ti}{I}   &1.54  $\pm$  0.12        &       27      \\
\ion{V}{I}        &1.69  $\pm$  0.30    &       4       \\
                
\ion{Cr}{I}   &0.95  $\pm$  0.10        &       17      \\
\ion{Mn}{I}   &1.03  $\pm$  0.01        &       7       \\
                                        
\ion{Fe}{I}   &0.96  $\pm$  0.11        &       164     \\
\ion{Ni}{I}   &0.97  $\pm$  0.10        &       10      \\
                                        
\ion{Sr}{II}  &1.04  $\pm$  0.06        &       3       \\
\hline
\end{tabular} 
\end{table}

\section{The line compendium} \label{Compendium}
With this paper we publish a complete list of 16\,431 
lines detected in the CRIRES-POP spectrum of 10 Leo, 91\% of which have the line-producing element or molecular species assigned.
Combining this list with the fully reduced spectrum from \citet{Nicholls_2017} available online, we have provided everything necessary for the production of a comprehensive infrared spectral atlas for 10 Leo.
The line lists are made available via machine-readable tables.

The basic structure of this first table is shown in Table \ref{table:YJ_list}.
A single-letter code in the rightmost column of the table distinguishes various levels of agreement with the VALD3 database and the Arcturus atlas, respectively.
The letter code is explained in Table \ref{R1}.
For instance, while lines in category A are found in the spectra of 10 Leo and Arcturus and in the VALD3 database, lines in category B are present in the 10 Leo spectrum and in VALD3, but are not listed in the Arcturus atlas.
In addition, we provide an indicator for the level of blending and quality for each line with the following meaning: S=single component, D=two components, T=three components, Q4, Q5,...= 4,5,... components, and X=uncertain continuum level affecting the shape and depth of the line.
The latter category accounts for about one-third of all lines.
In these cases, the EW given is affected by a higher uncertainty.

Categories C, D, and G indicate lines that are expected from the VALD3-based model or that have been identified in the Arcturus atlas, but do not have a counterpart in the 10 Leo spectrum.
We provide separate online tables for each of these cases using a structure similar to Table \ref{table:YJ_list}.
The lines can be broken down  into cases where the line is clearly not visible in 10 Leo ('absent'), and cases where a final decision on the presence is not possible due to noise or missing parts of the 10 Leo spectrum ('inconclusive').
Finally, category G also includes 28 lines where the identification in the Arcturus atlas and from VALD3 is not the same ('IDconflict'). 
The last row in Table \ref{R1} indicates the number of lines in each band without a known identification (category U). 
These are a subset of the total number of lines in each wavelength band given in the next-to-last row of the table, and are also a subset of the lines in category F.
The difference in number between F and U corresponds to the 396 lines we were able to newly assign to molecular and atomic species based on our computations of the expected locations of various, previously unidentified CO lines, and lines from the \citet{Peterson_2022} collection of Fe lines.

\begin{table*}
    \centering
        \caption{Excerpt from line list for the YJ band}        
        \label{table:YJ_list}
        \centering
\begin{tabular}{lccrrcccc}
\hline\hline
Wave&Absorptance&Equivalent&\multicolumn{2}{c}{Wavelength}&Species&Transition&Line &Catalogue\\
number&&width&observed&VALD3&&&quality and &Match\\
v(cm$^{-1}$)    & 1-F   & EWO(mA)       & WLobs(nm)     &WL (nm) & ID   &  &complexity    & \\
\hline
10387.73        &0.196  &43     &962.67 &962.68 &CN     &CN 1-0Q$\{$1$\}$50.5   &D    &B\\    
10387.21        &0.075  &4      &962.72 &962.72 &CN     &CN 2-1Q$\{$12$\}$25.5  &Q4     &B\\    
10387.15        &0.096  &20     &962.73 &962.73 &CN     &CN 2-1P$\{$1$\}$26.5   &Q4     &A\\    
10386.81        &0.103  &22     &962.76 &962.76 &CN     &CN 1-0R$\{$1$\}$59.5   &Q4     &A\\    
10386.12        &0.127  &19     &962.82 &962.83 &\ion{Fe}{I}&                   &D       &B\\    
10386.06        &0.107  &2      &962.83 &962.84 &\ion{Fe}{I}&                   &D       &B\\    
10385.71        &0.208  &33     &962.86 &962.87 &CN     &CN 1-0P$\{$2$\}$42.5   &X      &A\\    
10385.19        &0.290  &79     &962.91 &962.91 &\ion{Fe}{I}&                    &D      &B\\    
10384.85        &0.193  &41     &962.94 &962.95 &\ion{Fe}{I}&                    &D      &A\\    
10382.94        &0.102  &21     &963.12 &963.12 &CN     &CN 2-1R$\{$1$\}$43.5   &X      &A\\    
\hline
\end{tabular}
\begin{tablenotes}
\item{The full table is available at the CDS.}
\end{tablenotes}
\end{table*}

\begin{table*}
    \centering
        \caption{Number of line matches between 10 Leo (observed), VALD3 (calculated), and Arcturus (observed, Hinkle atlas)}        
        \label{R1}
\begin{tabular}{lcccccccccl}
\hline\hline
Category& 10 Leo & VALD3 & Arcturus &   YJ      &       H           &   K           &    L       &       M           &   sum      \\
\hline
A       & y & y & y &   1367    &       1375    &       971         &   358     &       1078    &       5149 \\
B       & y & y & n &   1941    &       2958    &       1984    &       824     &       1498    &       9205\\
C       & n & y & n &   1003    &       2104    &       1184    &       702     &       491         &    5484 \\
D       & n & y & y &   9           &   113         &   13          &   3   &     7           &   145      \\
E       & y & n & y &   60          &   84      &       13          &   93  &      1           &   251 \\
F       & y & n & n &   729     &       324         &   245     &       444     &       84         &       1826 \\
G       & n & n & y &   4           &   82          &   15          &   67  &      7           &   175      \\
\hline
sum &   &   &   &       5113    &       7040    &       4424    &       2491&   3166    &       22235\\
U       &       &   &   & 436       &   303     &       196         &   415     &       84         &       1434 \\
\hline
\end{tabular} 
\end{table*}

High-quality spectra of stars can be used to determine various spectral parameters for the CO molecule such as the rotational constants B, the ro-vibrational coupling constant $\alpha_e$, the centrifugal distortion constant, and the transition energies of the vibrational states of $^{12}$CO and $^{13}$CO.
Such results are complementary to laboratory measurements that are limited in, for example, temperature.
\citet{1991JMoSp.149..375F} used this approach to derive parameters of the CO molecule from the infrared solar spectrum.
The CRIRES-POP 10 Leo spectrum is of sufficient quality to perform a similar study. 
Resulting values are presented in Table \ref {table:10} together with the corresponding values from the literature. 
While there is a good agreement for most constants, we note some differences, most remarkably for $\omega_{e}$ for $^{13}$CO.

\begin{table*}
\centering{
\begin{threeparttable}
        \centering
        \caption{Overview of derived molecular constants for \element[][12]{C}O and \element[][13]{C}O (all quantities  are in units of cm\textsuperscript{-1})}        
        \label{table:10}
    \begin{tabular}{ccccc}
\hline\hline
&\multicolumn{2}{c}{\element[][12]{C}O}&\multicolumn{2}{c}{\element[][13]{C}O}\\
Parameter                                                       &This work  &Literature    &This work   &Literature\\
\hline
$\omega\textsubscript{e}$                       &2169.76    &2169.756 [1]  &2121.37               &2141.41 [2]\\
$\omega\textsubscript{0}$(1-0)          &2143.27    &2143.27 [3]   &2096.07         &2096.07 [3]\\
$\omega\textsubscript{0}$(2-0)          &4260.07    &4260.06 [3]   &4166.83         &4166.82 [3]\\
$\omega\textsubscript{0}$(3-0)          &6350.43    &6350.44 [3]   &                &6212.32 [3]\\
$\omega\textsubscript{0}$(3-1)          &4207.17    &4207.17 [3]   &4116.25         &4116.25 [3]\\
$\omega\textsubscript{0}$(4-2)          &4154.41    &4154.41 [3]   &4065.81         &4065.81 [3]\\
B\textsubscript{e}                              &1.9309     &1.9316 [1]    &1.8461       &                   \\
B\textsubscript{0}                              &1.9222     &1.9225 [3]    &1.8372               &1.8380 [3]     \\
B\textsubscript{1}                              &1.9049     &1.9050 [3]    &1.8215               &1.8216 [3]     \\
B\textsubscript{2}                              &1.8872     &1.8875 [3]    &1.8055               &1.8053 [3]     \\
B\textsubscript{3}                              &1.8700     &1.8700 [3]    &1.7890               &1.7890 [3]     \\
$\omega\textsubscript{e}$$\chi\textsubscript{e}$&13.236&13.288 [1] &12.640      &12.67 [2]     \\
$\alpha\textsubscript{e}$                       &1.7490E-02 &1.7505E-02 [1]&1.6351E-02  &                \\
D\textsubscript{e}                                      &6.120E-06      &6.122E-06 [1] &5.592E-06&5.593E-06 [3]    \\
r\textsubscript{e}(Å)                          &1.1242     &1.1282 [1]    &                     &                   \\
\hline
\end{tabular}
\tablebib{[1] \citet{Le_Floch_1991}; [2] \citet{Gendriesch_2009}; [3] \citet{Coxon_2004}}
\begin{tablenotes} 
\item\textbf{Parameters.} $\omega\textsubscript{e}$= harmonic wavenumber;
$\omega\textsubscript{0}$=$\Delta$E ($\nu$:1$\leftarrow$0); B\textsubscript{e}=equilibrium rotational constant, B\textsubscript{0}=rotational constant ($\nu$=0); $\omega\textsubscript{e}$$\chi\textsubscript{e}$=anharmonicity term; r\textsubscript{e}(Å)=bond distance; $\alpha\textsubscript{e}$=vibration-rotation interaction constant; D\textsubscript{e}=centrifugal distortion constant.
\end{tablenotes}
\end{threeparttable}}
\end{table*}

\section{Element abundances for 10 Leo and Arcturus}\label{sect:abundances}

A set of synthetic spectra has been compiled to derive elemental abundances for 10 Leo and Arcturus from the observed high-resolution NIR spectra. 
In a first step, a line catalogue with respective temperature, log \textit{g}, and microturbulence was requested from VALD3 based on solar abundances. 
We then computed a synthetic spectrum on the basis of the Model Atmospheres with Radiative and Convective Scheme (MARCS)  structure using the tool Synth3 \citep{Kochukhov_2012}.
Differences between the observed and the computed spectrum were analysed for individual elements. 
Based on this, element abundances in the VALD3 line catalogue file were modified for a new computation of the synthetic spectrum. This iterative process was repeated until a consistent set of abundances for the synthetic spectrum was obtained. 
To  find the best fit, we both minimised the differences for EWs and carried out a visual inspection of an overlay of the observed and synthetic spectrum.
Furthermore, the computed and observed EWs of selected lines were compared in unclear cases individually.
\begin{table}
    \begin{center}
        \caption{Abundances for 10 Leo and Arcturus derived from spectral synthesis. Column 2:  solar abundances scaled to the metallicity of 10 Leo. Column 3: abundances for 10 Leo derived in this paper by spectral synthesis. Column 4: abundances for Arcturus derived in this paper. Column 5: differences between 10 Leo and Arcturus abundances.}   
        \label{R21}
\begin{tabular}{ccccc}
\hline\hline
{Element\textsuperscript{a}}&\multicolumn{4}{c}{Abundances}\\
    &\bf{Solar scaled\textsuperscript{b}}       &       10 Leo  & Arcturus & Leo-Arc   \\
\hline
C       &       -3.52   &       -3.85$\pm{0.05}$        &       -4.14   &       0.29    \\              
N       &       -4.12   &       -4.05$\pm{0.09}$        &       -4.37   &       0.32    \\              
O       &       -3.21   &       -3.35$\pm{0.09}$        &       -3.27   &       -0.08   \\
Na      &       -5.71   &       -5.80$\pm{0.02}$        &       -5.89   &       0.09    \\
Mg      &       -4.46   &       -4.46$\pm{0.17}$        &       -4.48   &       0.02    \\
Al      &       -5.57   &       -5.40$\pm{0.08}$        &       -5.41   &       0.01    \\
Si      &       -4.49   &       -4.80$\pm{0.07}$        &       -4.83   &       0.03    \\
K       &       -6.92   &       -7.00$\pm{0.09}$        &       -7.09   &       0.09    \\              
Ca      &       -5.68   &       -6.00$\pm{0.06}$        &       -6.08   &       0.08    \\
Sc      &       -8.87   &       -9.00$\pm{0.52}$        &       -9.10   &       0.10    \\
Ti      &       -7.02   &       -7.30$\pm{0.13}$        &       -7.39   &       0.09    \\
V       &       -8.04   &       -8.30$\pm{0.31}$        &       -8.74   &       0.44    \\
Cr      &       -6.37   &       -6.55$\pm{0.07}$        &       -6.67   &       0.12    \\
Mn      &       -6.65   &       -6.00$\pm{0.10}$        &       -6.18   &       0.18    \\              
Fe      &       -4.54   &       -4.80$\pm{0.17}$        &       -4.96   &       0.16    \\
Ni      &       -5.79   &       -6.25$\pm{0.08}$        &       -6.40   &       0.15    \\              
Sr      &       -9.07   &       -9.00$\pm{0.05}$        &       -9.09   &       0.09    \\
\hline
\end{tabular}
\end{center}
\textsuperscript{\bf{a}}Only neutral lines were used for all elements except for Sr, for which only ionised lines were used.
\textsuperscript{\bf{b}}The solar-scaled abundances are based on the element abundances in the solar photosphere derived by \citet{Grevesse_1998}.
\end{table}

\begin{figure*}
        \sidecaption
        \includegraphics[width=12cm]{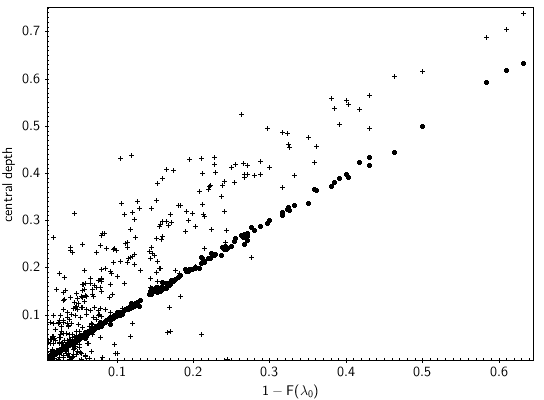}  
                \caption{cdepth deduced from the Gaussian fit versus \ion{Fe}{I} lines (1-F\textsubscript{($\lambda\textsubscript{0}$)}). The dots are from individual Gaussian fits to isolated Fe lines in the observed 10 Leo spectrum. The crosses refer to cdepths from the VALD3 extracted stellar calculation for 10 Leo based on solar Fe abundances.}
        \label{figure:1}
\end{figure*}

Fig. \ref{figure:1} shows a comparison of the observed and computed central depths for about 340 isolated \ion{Fe}{I} lines selected from across the range of the 10 Leo spectrum. 
Each observed iron line was fitted by a Gaussian curve to derive the respective position, central depth, full width at half maximum (FWHM), and the EW. 
The central depth from this fit is compared with the observed value of 1-F, where F is the normalised flux at the line centre $\lambda\textsubscript{0}$ (filled circles). 
The second dataset in this plot (crosses) refers to the same lines with the central depth now derived from the VALD3 {\tt stellar extraction} model for solar abundances (see Sect.\,\ref{line_id}) with the observed value of 1-F. 
The Gaussian model  provides a good fit to the observed line profiles over the whole range of line strengths studied here.
The comparison between the observed and the modelled central depths shows a systematic difference in line depth and  considerable scatter around the relation.
The systematic difference indicates that 10 Leo has a lower iron abundance than the Sun.
The high S/N of the spectrum and the good fit of the observed lines by a Gaussian profile, which indicates a low effect of line blending, suggests that the scatter we see between observed and modelled line depths is only partly due to observational noise, but significantly due to uncertainties in the theoretical line data.
Indeed, a number of \ion{Fe}{I} lines predicted by the synthetic spectrum could not be observed in the 10 Leo spectrum as discussed in Sect.\,\ref{Unidentified lines} (Tables \ref{table:8}, \ref{table:13}, \ref{table:14}, and \ref{table:15}).

In total, abundances for 17 key elements were determined from the 10 Leo and Arcturus spectra. 
The third and fourth column of Table \ref{R21} reflect the set of abundances for the best fit for 10 Leo and Arcturus, respectively, whereas the second column lists the solar  reference values used as the starting point.
Abundances for C, N, and O were derived from molecular lines using the approach described in \citet{Hinkle_etal_2020}. In principle, the abundances of these three elements were altered in each fitting step until a simultaneous fit of the lines of CO, CN, and OH was achieved.
The fifth column shows the abundance differences between the two K giants.
Abundance uncertainties were derived from the standard deviation of the ratios in EWs between the observed and the synthetic spectrum, and are given in column three of Table \ref{R21} for 10 Leo.
For Arcturus, the listed abundances come with similar uncertainties.
As a K giant, 10 Leo should show an abundance pattern typical for a post first dredge-up object.
According to evolutionary models \citep{Karakas_Dawes_Review, 2015ApJS..219...40C}, we thus expect a reduction of $^{12}$C and an enhancement of $^{14}$N on the stellar surface. 
Both trends are clearly visible in Table \ref{R21}.
The C and O abundances are in excellent agreement with the values derived by \citet{2015AJ....150...88L} from optical spectra. 
The same authors give abundances for various other elements in 10 Leo, most of them indicating an overabundance compared to the Sun.
From our analysis, however, we find most elements to be slightly underabundant compared to the Sun.
Solar-like abundances for 10 Leo were reported by \citet{2015A&A...580A..24D}.
While error bars both in our analysis and for the literature values do not allow a conclusive comparison, we note that our analysis tends to give a lower mean metallicity for 10 Leo.
\citet{2016ApJ...823...36T} independently derived Si abundances from the CRIRES-POP spectrum of 10 Leo and got [Si/Fe]=-0.04$\pm{0.03}$, which is in satisfying agreement with our findings.

Using second overtone lines of $^{12}$CO and $^{13}$CO, we derived a $^{12}$C/$^{13}$C ratio of 22$\pm{8}$ using a curve of growth analysis \citep{2016ApJ...825...38H}.
To our knowledge, the carbon isotopic ratio had not been measured for this star before.
To derive ratios for the main isotopes of oxygen, we used fundamental lines of C$^{16}$O, C$^{17}$O, and C$^{18}$O in the M band. We find $^{16}$O/$^{17}$O = 299$^{-107}_{+241}$ and $^{16}$O/$^{18}$O = 345$^{-125}_{+276}$.
These values agree with model expectations for a 2\,M$_{\sun}$ red giant after a first dredge-up and observations for similar stars \citep{2015A&A...578A..33L, 2015ApJS..219...40C}.

To estimate the reliability of our abundance determinations for 10 Leo in the absence of a larger set of literature data, we can compare our results for Arcturus, which have been derived in an identical manner, with literature results. We thereby focus on the most recent papers.
Very recently measured abundances of Na, Mg, and Al by \citet{Lind_Arcturus} agree within the error bars with our findings. That paper also gives abundance corrections for non-local thermodynamic equilibrium (NLTE), which have not been taken into account in our analysis.
\citet{Fukue_Arcturus} present abundances for C, N, O, Mg, Si, Ca, Ti, Cr, Fe, and Ni, all of which coincide with our findings within 0.1 dex, except for Cr, where \citet{Fukue_Arcturus} found an abundance 0.33 dex lower than our result. Further abundances of key elements were presented by \citet{Joensson_Arcturus} and \citet{Lomaeva_Arcturus}, partly based on the spectrum from the Arcturus atlas.
Results for Mg, Ca, Ti, Sc, V, and Ni agree within our error bars.
Their O abundance is slightly lower than our value.
The abundances from the two mentioned papers coincide with minor differences with the major study by \citet{Jofre_Arcturus}. The Cr abundance from \citet{Lomaeva_Arcturus} agrees with the results from \citet{Fukue_Arcturus} and thus significantly differs from our findings. The reason for this difference is not clear. In addition, our Mn abundance is much larger than the value presented in \citet{Lomaeva_Arcturus} and in \citet{Jofre_Arcturus}. Since we also see an outstanding overabundance of this element in 10 Leo compared to the sun, we have to assume the Mn abundances derived in this paper to be potentially erroneous.  
We aim to understand this discrepancy in a separate study beyond the scope of this paper, which is presenting the line compendium in the CRIRES-POP spectrum of 10 Leo.
Except for these two elements, abundances derived for 10 Leo and Arcturus in this paper coincide well with other studies.

\section{Summary}
The main goal of this paper was to provide a compendium of line identifications for a K giant that is as complete as possible.
This includes more than 16\,000 lines from molecules and atoms, detected in the pipeline-reduced and fully corrected high-resolution NIR spectrum of the cool star 10 Leo, and their assignment to atomic and molecular species. 
These results are important for computing synthetic spectra, and abundance determination for evolved stars in general. 
In addition, the search for lines from neutron capture elements profits from a high degree of completeness and correctness of line lists for Fe and below, including molecular lines. This can assist in determining interfering lines that may completely mask or mimic lines of possible neutron capture elements.

The identification of absorption lines has been done visually with the help of various line lists, {\tt stellar extraction} from VALD3, and other published sources. 
Hydrogen and lines from various further elements were identified, including C, Na, Mg, Al, Si, P, S, K, Ca, Sc, Ti, V, Cr, Mn, Fe, Co, Ni, Cu, Zn, Rb, Sr, Y, Zr, Ce, and Dy. 
Some of these elements, for example Sr, Ce, and Dy, are only detectable in their ionised form. 

Molecules such as CO, CN, and OH dominate the spectrum. Absorption lines from \element[][13]{C}O, C\element[][17]{O,} and C\element[][18]{O} were easily identified in several bands and with sufficient strengths to determine isotopic ratios. 
In contrast, absorption lines from \element[][13]{C}N are very weak and only identified for the 0-0 bands. 
Lines of C\element[][15]{N} are not detectable in the spectrum.

The comparison between observed and synthetic spectra allowed us to identify a large number of lines that are present in the synthetic spectra but are not observed, and lines that are observed and identified, but not present in the synthetic spectra. 
The high number of discrepancies reveals the need to improve the spectroscopic data for metals, for example, for Mg and Fe.  
The $L$ band shows the largest fraction of unidentified lines (29\%), and the $M$ band the lowest (6\%). Lists of unidentified and missing lines are also part of our line compendium.
The quality of the spectrum allows for various spectral parameters for CO to be tested such as the rotational constant, vibrational coupling constant, distortion constant, and  transition energies of the vibrational states. 
We find a good agreement with literature values for $^{12}$C$^{16}$O, but some discrepancies for $^{13}$C$^{16}$O, which require further investigation.

Finally, we have presented abundances for 17 key elements from the CRIRES-POP spectrum of 10 Leo and the FTS spectrum of Arcturus, which are in good agreement with literature values in most cases. 
This paper complements the CRIRES-POP spectrum of 10 Leo published by \citet{Nicholls_2017} providing a complete NIR atlas of this K giant.
In comparing it with the Arcturus atlas, metallicity and temperature effects can be studied for individual sections of the 1 to 5\,$\mu$m range.

\begin{acknowledgements}
This work has made use of the VALD3 database, operated at Uppsala University, the Institute of Astronomy RAS in Moscow, and the University of Vienna. The authors wish to thank Bernhard Aringer and Kenneth Hinkle for valuable suggestions and discussions on the paper.
\end{acknowledgements}

\bibliographystyle{aa} 
\bibliography{references} 

\begin{appendix}

\section{Unassigned and unobserved lines}
This appendix presents tables of lines present in the 10 Leo spectrum but not assigned to an element or molecule yet (Tables \ref{table:11} and \ref{table:12}). 
In addition, we give tables of lines expected to be visible from the VALD3 database, but not observed in the spectra of 10 Leo or Arcturus (Tables \ref{table:13} to \ref{table:15}).
For the YJ band, the corresponding tables can be found in the main part of the paper (Tables \ref{table:7} and \ref{table:8}).

\begin{table*}
\caption{Most intense unassigned lines with an EW>100 mÅ in the H band and possible contributors}
\label{table:11}
\begin{tabular}{lcll}
\hline\hline
wavelength & EW & possible blends & reference\\
(nm)      &mÅ&               &                  \\
\hline  
1415.75 &       122     &       \ion{Fe}{II} 1415.74 \ion{Ti}{II} 1415.75, \ion{Ni}{II} 1415.76 \ion{Cs}{I} 1415.78, \ion{Cr}{II} 1415.81 &        (1)         \\
1416.90 &       161     &       \ion{Co}{I} 1416.90, \ion{Mn}{II} 1416.94, \ion{V}{II} 1416.92     &       (1)         \\
1418.00 &       186     &               &                   \\
1418.18 &       187     &       1418.18 CN 0-1Q(2)13.5, \ion{Fe}{II} 1418.17, \ion{Ni}{I} 1418.17, \ion{Mn}{i} 1418.18,&\\
&&\ion{V}{i} 1418.18, \ion{Sc}{II} 1418.18      &       (1)         \\
1421.09 &       102     &                   \\
1421.54 &       146     &       \ion{Fe}{II} 1421.52, \ion{Sc}{I} 1421.51, \ion{Mn}{II} 1421.55, \ion{Cr}{I} 1421.56       &       (1)         \\
1424.87 &       177     &       \ion{Fe}{I} 1421.90     &       (1)         \\
1425.29 &       316     &       \ion{Fe}{I} 1425.30 (CN0-1R(2)31.5 subtracted)  &                   \\
1426.37 &       185     &       \ion{Fe}{I} 1426.39, \ion{Ca}{I} 1426.34        &       (1)         \\
1426.81 &       134     &       \ion{Ti}{I} 1426.81, \ion{Fe}{I} 1426.89        &       (1)         \\
1427.65 &       151     &       \ion{Fe}{II} 1427.68, \ion{P}{I} 1427.67, \ion{C}{I} 1427.64, \ion{Ni}{I} 1427.65 &       (1)         \\
1427.67 &   121 &               &\\
1427.75 &       114     &       \ion{Mn}{I} 1427.74, \ion{Fe}{I} 1427.74        &       (1)         \\
1428.38 &       222     &       \ion{V}{I} 1428.39, 1428.40     &       (1)         \\
1429.56 &       142     &       \ion{Mn}{II} 1429.51, (CN 0-1R(1)33.5 subtracted)       &       (1)         \\
1429.88 &       103     &       \ion{Mn}{I} 1429.87     &       (1)         \\
1429.90 &       146     &       \ion{Sc}{II} 1429.90, \ion{V}{II} 1429.91       &       (1)         \\
1432.89 &       102     &       \ion{Mn}{I} 1432.88, \ion{Fe}{I} 1432.84, \ion{Fe}{II} 1432.85    &       (1)         \\
1434.09 &       143     &       \ion{Fe}{II} 1434.09, \ion{Ti}{I} 1434.09, \ion{Si}{II} 1434.07    &       (1)         \\
1449.79 &       258     &       \ion{Ti}{I} 1449.79, \ion{Fe}{II} 1449.78, \ion{Mn}{II} 1449.74,&\\
&&\ion{V}{i} 1449.76, \ion{C}{I} 1449.77, \ion{V}{I} 1449.83    &       (1)         \\
1452.93 &       230     &       \ion{Fe}{I} 1452.93, \ion{Mn}{I} 1452.92,\ion{Sc}{I} 1452.94         &       (1)         \\
1486.66 &       158     &       \ion{Fe}{II} 1486.61, \ion{Si}{I} 1486.69       &       (1)         \\
1581.45 &       142     &       \ion{Fe}{I} 1581.45(cdepth 0.037)       &       (1),(2)     \\
1597.20 &       251     &       \ion{Fe}{I} 1597.20, \ion{Al}{I} triplet 1597.27 (cdepth 0.288)  &       (1),(2)     \\
1597.73 &       129     &       \ion{V}{I} 1597.77      &       (1)         \\
1608.03 &       187     &       \ion{Ti}{I} 1607.98, \ion{Fe}{I} 1608.03 (cdepth 0.048)  &       (1),(2)     \\
1643.34 &       135     &       \ion{Ni}{I} 1643.34, \ion{Fe}{II} 1643.30       &       (1)         \\
1696.63 &       143     &       \ion{Fe}{I} 1696.65 (cdepth 0.090, \ion{Cr}{I} 1696.66, \ion{Ti}{I} 1696.97    &       (1)         \\
1707.21 &       113     &       \ion{Fe}{I} 1707.22 (cdepth 0.055), \ion{Fe}{II} 1707.22 &       (1),(2)     \\
1758.51 &       164     &       \ion{Co}{I} 1758.47, \ion{Cr}{II} 1758.46, \ion{Fe}{I} 1758.53     &       (1),(2)     \\
1774.44 &       189     &               &                   \\
1775.22 &       105     &       \ion{Fe}{I} 1775.22 (cdepth 0.060)&     (1)         \\
1794.36 &       154     &       1794.40 \ion{Cr}{II} 1794.33, \ion{P}{I} 1794.34, \ion{Si}{I} 1794.38, \ion{K}{I}    &(1)            \\

\hline
\end{tabular}
\tablebib{(1)\citet{Ryabchikova_2015}, (2)\citet{Hinkle_1995}}
\end{table*}

\begin{table*}
\begin{threeparttable}
\caption{Most intense unassigned lines with an EW>100 mÅ in the K, L, and M band and possible contributors}
\label{table:12}
\begin{tabular}{l c l c}
\hline\hline
wavelength&EW   &possible blends                        &reference\tnote{a} \\
(nm)       &mÅ  &                                       &                   \\
\hline
1978.17 &   241 & strong shoulder on \ion{Ca}{I} 1978.22&      \\
1979.92 &       163     &       \ion{Mn}{I} 1979.90, cdepth 0.039       &       (1)\\
1986.70 &   150 & strong shoulder on \ion{Ca}{I} 1986.76&      \\
2009.53 &       213     &       \ion{Co}{II} 2009.55, \ion{Fe}{II} 2009.58,
\ion{Mn}{i} 2009.48     &(1)\\
2106.54 &       265     &       \ion{Cr}{I} 2106.56                     &       (1)\\
2163.17 &       112     &       \ion{Ti}{I} 2163.24, \ion{Fe}{II} 2163.25,
\ion{V}{i} 2163.12, \ion{Ca}{I} 2163.11 &       (1)         \\
2478.67 &       157     &                                               &      \\
2499.27 &       189     &       \ion{Al}{I} 2499.28, cdepth 0.073       &       (1)\\
2508.83 &       224     &       \ion{Ni}{II} 2508.82                &   (1)\\
3415.38 &       381     &       \ion{Ti}{I} 3415.36, \ion{Cr}{I} 3415.41&(1)\\
3416.00 &       410     &       \ion{Cr}{I} 3416.04                     &       (1)\\
3542.12 &       104     &       \ion{Fe}{II} 3542.11                &   (1)\\
3544.45 &       103     &       \ion{V}{I} 3544.45                      &       (1)\\
3546.79 &       168     &       \ion{Fe}{II} 3546.77                &   (1)\\
3983.61 &   109 &                                       &      \\
3996.26 &       103     &       doublet ( 3996.13 /\ion{Co}{I} 3996.23)&    \\
3999.95 &       127     &       \ion{C}{I} 3999.94, \ion{Cr}{II} 4000.00&(1)\\
4006.27 &       103     &                                               &       \\
4160.42 &       104     &       \ion{Fe}{I} 4160.31                     &       (2) \\
4167.66 &       127     &                                               &       \\
4549.36 &   393 & wavelength correction of -0.11 nm applied& \\
4553.23 &   188 & wavelength correction of -0.11 nm applied& \\
4565.37 &       393     & wavelength correction of -0.08  nm applied,   \element[][13]{C}O 2-1R40, C\element[][18]{O} 2-1R42,&\\ 
        &       &C\element[][18]{O} 3-2R27 subtracted&       \\
4569.88 &       426     & wavelength correction of -0.03  nm applied,   \element[][13]{C}O 2-1R39, C\element[][17]{O} 2-1R80  subtracted&    \\
4572.17 &       362     &       C\element[][17]{O} 3-2R71, 
C\element[][18]{O} 4-3R85 subtracted    &    \\
4574.46 &       644     &       CO 2-1R38 subtracted                &        \\
4576.77 &       449     &       Co1 4576.76 \element[][13]{C}O 4-3R74,
\element[][13]{C}O 2-1R38 subtracted    &       (1)         \\
4579.11 &       356     &                                               &         \\
4580.28 &       472     &       \ion{Fe}{II} 4580.31, quartet:  CO 1-0R10,\element[][13]{C}O 3-2R50, C\element[][18]{O} 4-3R76, 
CO 4-3R39 subtracted            &   (1)             \\
4581.43 &       190     &       \ion{Fe}{I} 4581.53, 
\element[][13]{C}O 4-3R70, CO subtracted                    &(1) \\
4583.80 &       109     &                                               &          \\
4595.01 &       168     &       \ion{Ti}{I} 4594.94                     &       (1)        \\
4595.86 &       146     &        CO 3-2R25, CO 3-2R46 subtracted        &              \\
4625.85 &       640     &                                               &                           \\
4678.84 &       416     &       \element[][13]{C}O 4-3R40 subtracted&          \\
4682.45 &       446     &       \element[][13]{C}O 6-5R85 subtracted&          \\
4705.95 &       299     &                                               &          \\
4726.57 &       721     &       CO 1-0P7 subtracted                     &          \\
4748.62 &       211     &                                               &          \\
4794.91 &       210     &       C\element[][18]{O} 3-2R12 subtracted&              \\
4798.53 &       273     &       \ion{Fe}{I} 4798.55 subtracted          &          \\
4822.08 &       177     &       \element[][13]{C}O 6-5R35 subtracted&          \\
4874.71 &       110     &       \element[][13]{C}O 9-8R80 subtracted&          \\
4882.62 &       230     &       C\element[][17]{O} 1-0P17, CO 10-9R110, \element[][13]{C}O 8-7R49 subtracted       & \\
4883.87 &       500     &       CO 10-9R57, \element[][13]{C}O 7-6R35, 
\element[][13]{C}O 9-8R72 subtracted    &                   \\
5023.05 &       130     & blended with \element[][13]{C}O 4-3P8, 
CO 12-11R56& \\
5087.31 &       133     &                                               &          \\
5186.19 &       122     &       \ion{Mn}{I} 5186.29                     &       (1)    \\
\hline
\end{tabular}
\tablebib{(1)\citet{Ryabchikova_2015}, (2)\citet{Hinkle_1995}}
\end{threeparttable}
\end{table*}

\begin{table*}
\caption{Lines with cdepth>0.10 not present in the observed H-band spectrum.}
\label{table:13}
\begin{tabular}{l c c c | l c c c}
\hline\hline
Wavelength&species&Absorptance\tnote(\textsuperscript{a}&Synthetic\tnote(\textsuperscript{b}&Wavelength&species&Absorptance\tnote(\textsuperscript{a}&Synthetic\tnote(\textsuperscript{b}\\ 
(nm)&ID  &1-F        &   cdepth     &  nm&ID  &1-F        &   cdepth     \\
\hline
1424.28 &       \ion{Fe}{I}     &       0.004   &       0.277   &       1650.86 &       \ion{Fe}{I}     &       0.015   &       0.293   \\
1429.42 &       \ion{Fe}{I}     &       0.001   &       0.268   &       1655.23 &       \ion{Fe}{I}     &       -0.008  &       0.115   \\
1432.78 &       \ion{Fe}{I}     &       -0.006  &       0.156   &       1656.12 &       \ion{Fe}{I}     &       0.058   &       0.113   \\
1449.58 &       \ion{Fe}{I}     &       0.008   &       0.283   &       1658.59 &       \ion{Fe}{I}     &       0.015   &       0.392   \\
1460.77 &       \ion{Fe}{I}     &       0.006   &       0.140   &       1659.43 &       \ion{Si}{I}     &       0.007   &       0.133   \\
1461.02 &       \ion{Mg}{I}     &       -0.030  &       0.120   &       1659.77 &       \ion{Ni}{I}     &       0.023   &       0.103   \\
1465.34 &       \ion{Mg}{I}     &       0.088   &       0.166   &       1659.78 &       \ion{S}{I}      &       0.013   &       0.108   \\
1465.35 &       \ion{Mg}{I}     &       -0.014  &       0.119   &       1663.44 &       \ion{Fe}{I}     &       0.016   &       0.247   \\
1481.13 &       \ion{Mg}{I}     &       0.012   &       0.123   &       1664.52 &       \ion{Fe}{I}     &       0.025   &       0.162   \\
1486.77 &       \ion{Fe}{I}     &       0.015   &       0.129   &       1664.98 &       \ion{Fe}{I}     &       0.000   &       0.115   \\
1487.07 &       \ion{Fe}{I}     &       0.023   &       0.203   &       1665.18 &       \ion{Fe}{I}     &       -0.002  &       0.233   \\
1494.12 &       \ion{Fe}{I}     &       -0.001  &       0.227   &       1681.52 &       \ion{Fe}{I}     &       -0.005  &       0.103   \\
1501.35 &       \ion{Fe}{I}     &       -0.001  &       0.261   &       1682.26 &       \ion{Fe}{I}     &       0.035   &       0.180   \\
1511.03 &       \ion{Fe}{I}     &       0.090   &       0.129   &       1682.62 &       \ion{Fe}{I}     &       -0.013  &       0.106   \\
1515.94 &       \ion{Fe}{I}     &       -0.013  &       0.242   &       1684.85 &       \ion{Fe}{I}     &       0.052   &       0.258   \\
1519.09 &       \ion{Ti}{I}     &       -0.018  &       0.156   &       1685.77 &       \ion{Ni}{I}     &       0.008   &       0.189   \\
1519.94 &       \ion{Ni}{I}     &       -0.017  &       0.142   &       1685.81 &       \ion{Fe}{I}     &       -0.015  &       0.179   \\
1521.83 &       \ion{Fe}{I}     &       -0.015  &       0.103   &       1686.31 &       \ion{Fe}{I}     &       0.011   &       0.455   \\
1527.57 &       \ion{Fe}{I}     &       -0.008  &       0.173   &       1688.32 &       \ion{Fe}{I}     &       0.016   &       0.225   \\
1544.86 &       \ion{Fe}{I}     &       -0.004  &       0.366   &       1688.82 &       \ion{Fe}{I}     &       0.004   &       0.175   \\
1544.86 &       \ion{Fe}{I}     &       0.000   &       0.285   &       1700.13 &       \ion{Co}{I}     &       0.075   &       0.317   \\
1548.01 &       \ion{Fe}{I}     &       0.048   &       0.218   &       1700.93 &       \ion{Fe}{I}     &       -0.034  &       0.143   \\
1548.86 &       \ion{Fe}{I}     &       -0.002  &       0.146   &       1701.50 &       \ion{Mg}{I}     &       0.018   &       0.120   \\
1549.03 &       \ion{Fe}{I}     &       0.008   &       0.274   &       1701.74 &       \ion{Fe}{I}     &       -0.010  &       0.289   \\
1551.49 &       \ion{Fe}{I}     &       -0.004  &       0.344   &       1702.98 &       \ion{Fe}{I}     &       0.009   &       0.244   \\
1559.12 &       \ion{Fe}{I}     &       0.005   &       0.201   &       1704.48 &       \ion{Fe}{I}     &       -0.004  &       0.219   \\
1562.20 &       \ion{Fe}{I}     &       0.026   &       0.186   &       1705.23 &       \ion{Fe}{I}     &       0.017   &       0.115   \\
1571.26 &       \ion{Ca}{I}     &       0.061   &       0.136   &       1705.75 &       \ion{Fe}{I}     &       -0.010  &       0.184   \\
1578.60 &       \ion{Fe}{I}     &       -0.013  &       0.142   &       1717.19 &       \ion{Fe}{I}     &       -0.012  &       0.130   \\
1579.38 &       \ion{Fe}{I}     &       0.028   &       0.424   &       1720.38 &       \ion{Fe}{I}     &       0.012   &       0.125   \\
1581.24 &       \ion{Si}{I}     &       0.001   &       0.189   &       1728.55 &       \ion{Fe}{I}     &       -0.002  &       0.138   \\
1585.84 &       \ion{Fe}{I}     &       0.037   &       0.246   &       1737.59 &       \ion{Fe}{I}     &       -0.003  &       0.254   \\
1590.32 &       \ion{Fe}{I}     &       0.057   &       0.293   &       1741.17 &       \ion{Ti}{I}     &       0.191   &       0.140   \\
1592.68 &       \ion{Fe}{I}     &       0.107   &       0.243   &       1741.21 &       \ion{Ni}{I}     &       0.298   &       0.103   \\
1592.98 &       \ion{Fe}{I}     &       -0.002  &       0.110   &       1745.73 &       \ion{Ca}{I}     &       -0.003  &       0.143   \\
1593.30 &       \ion{Fe}{I}     &       0.027   &       0.327   &       1746.39 &       \ion{Fe}{I}     &       -0.004  &       0.236   \\
1593.42 &       \ion{Fe}{I}     &       0.042   &       0.357   &       1747.24 &       \ion{Fe}{I}     &       0.075   &       0.327   \\
1593.65 &       \ion{Fe}{I}     &       0.003   &       0.405   &       1747.25 &       \ion{Fe}{I}     &       0.049   &       0.119   \\
1596.32 &       \ion{Mn}{I}     &       0.030   &       0.163   &       1750.52 &       \ion{Fe}{I}     &       0.019   &       0.229   \\
1596.80 &       \ion{Fe}{I}     &       0.012   &       0.174   &       1756.84 &       \ion{Fe}{I}     &       -0.033  &       0.142   \\
1596.95 &       \ion{Mn}{I}     &       0.127   &       0.212   &       1758.94 &       \ion{Fe}{I}     &       0.007   &       0.294   \\
1597.27 &       \ion{Al}{I}     &       0.033   &       0.139   &       1763.07 &       \ion{Fe}{I}     &       -0.057  &       0.233   \\
1600.66 &       \ion{Fe}{I}     &       0.019   &       0.213   &       1769.93 &       \ion{Fe}{I}     &       -0.005  &       0.379   \\
1600.78 &       \ion{Fe}{I}     &       0.022   &       0.171   &       1771.42 &       \ion{Fe}{I}     &       0.006   &       0.373   \\
1603.01 &       \ion{Fe}{I}     &       0.017   &       0.277   &       1775.37 &       \ion{Co}{I}     &       -0.002  &       0.118   \\
1606.18 &       \ion{Fe}{I}     &       0.019   &       0.160   &       1776.51 &       \ion{Fe}{I}     &       -0.028  &       0.312   \\
1606.91 &       \ion{Fe}{I}     &       0.015   &       0.261   &       1781.67 &       \ion{Fe}{I}     &       -0.024  &       0.133   \\
1608.12 &       \ion{Fe}{I}     &       -0.002  &       0.133   &       1783.24 &       \ion{Fe}{I}     &       0.008   &       0.176   \\
1611.55 &       \ion{Fe}{I}     &       0.016   &       0.101   &       1785.08 &       \ion{Fe}{I}     &       0.200   &       0.416   \\
1620.68 &       \ion{Fe}{I}     &       -0.001  &       0.229   &       1785.53 &       \ion{Fe}{I}     &       0.010   &       0.268   \\
1621.38 &       \ion{Fe}{I}     &       0.031   &       0.102   &       1786.53 &       \ion{Fe}{I}     &       -0.020  &       0.410   \\
1623.04 &       \ion{Fe}{I}     &       0.100   &       0.413   &       1786.56 &       \ion{Fe}{I}     &       0.010   &       0.132   \\
1623.70 &       \ion{Fe}{I}     &       0.035   &       0.210   &       1786.92 &       \ion{Fe}{I}     &       0.035   &       0.193   \\
1624.34 &       \ion{Fe}{I}     &       -0.008  &       0.255   &       1795.42 &       \ion{Ni}{I}     &       -0.018  &       0.124   \\
1628.52 &       \ion{Fe}{I}     &       0.042   &       0.101   &       1805.47 &       \ion{Y}{I}      &       -0.059  &       0.155   \\
1649.43 &       \ion{Fe}{I}     &       0.001   &       0.295   &       1807.11 &       \ion{Fe}{I}     &       0.006   &       0.136   \\
1649.45 &       \ion{Mn}{I}     &       -0.004  &       0.273   &                       &                       &               &               \\
\hline
\end{tabular}
\begin{tablenotes}
\item\textsuperscript{a}observed (1-F) at wavelength  
\item\textsuperscript{b}VALD3 stellar scaled to solar values
\end{tablenotes}
\end{table*}

\begin{table}
\caption{Lines with cdepth>0.10 not present in the observed K-band spectrum.}            
\label{table:14}    
\begin{tabular}{c c c c}     
\hline\hline                    
Wavelength&species&Absorptance\tnote(\textsuperscript{a}&Synthetic\tnote(\textsuperscript{b} \\ 
(nm)            &ID                     &1-F             &   cdepth   \\
\hline
1960.19 &       \ion{Ni}{I}     &       0.005   &       0.111   \\
2006.01 &       \ion{Cr}{I}     &       0.037   &       0.137   \\
2007.77 &       \ion{Fe}{I}     &       -0.006  &       0.223   \\
2009.19 &       \ion{Fe}{I}     &       -0.005  &       0.118   \\
2034.51 &       \ion{Fe}{I}     &       0.000   &       0.102   \\
2050.11 &       \ion{Ca}{I}     &       -0.006  &       0.168   \\
2052.72 &       \ion{Mg}{I}     &       -0.005  &       0.356   \\
2052.72 &       \ion{Mg}{I}     &       -0.003  &       0.107   \\
2055.31 &       \ion{Fe}{I}     &       0.015   &       0.104   \\
2083.06 &       \ion{Fe}{I}     &       0.004   &       0.113   \\
2083.91 &       \ion{Mg}{I}     &       0.085   &       0.337   \\
2087.28 &       \ion{Mg}{I}     &       0.000   &       0.202   \\
2123.95 &       \ion{Mg}{I}     &       0.001   &       0.323   \\
2152.80 &       \ion{Fe}{I}     &       0.011   &       0.145   \\
2176.70 &       \ion{Mg}{I}     &       0.010   &       0.310   \\
2195.73 &       \ion{Fe}{I}     &       0.000   &       0.157   \\
2257.78 &       \ion{Fe}{I}     &       -0.006  &       0.148   \\
2287.88 &       \ion{Fe}{I}     &       0.004   &       0.225   \\
2323.30 &       \ion{Fe}{I}     &       -0.001  &       0.160   \\
2329.29 &       \ion{Fe}{I}     &       0.013   &       0.200   \\
2350.59 &       \ion{Mg}{I}     &       0.005   &       0.137   \\
2351.67 &       \ion{Fe}{I}     &       0.040   &       0.247   \\
2369.02 &       \ion{Fe}{I}     &       -0.011  &       0.335   \\
2374.31 &       \ion{Mg}{I}     &       0.042   &       0.354   \\
2374.31 &       \ion{Mg}{I}     &       0.079   &       0.110   \\
2374.33 &       \ion{Mg}{I}     &       0.040   &       0.306   \\
2374.33 &       \ion{Mg}{I}     &       0.016   &       0.117   \\
2374.33 &       \ion{Mg}{I}     &       -0.005  &       0.104   \\
2374.34 &       \ion{Mg}{I}     &       -0.016  &       0.304   \\
2384.32 &       \ion{Mg}{I}     &       0.109   &       0.288   \\
2389.61 &       \ion{Si}{I}     &       0.019   &       0.124   \\
2394.56 &       \ion{Mn}{I}     &       0.003   &       0.121   \\
2415.14 &       \ion{Fe}{I}     &       0.028   &       0.151   \\
2417.00 &       \ion{Fe}{I}     &       0.194   &       0.225   \\
2425.13 &       \ion{Si}{I}     &       -0.008  &       0.178   \\
2438.05 &       \ion{Fe}{I}     &       0.092   &       0.212   \\
2453.29 &       \ion{Fe}{I}     &       0.023   &       0.263   \\
2453.93 &       \ion{Si}{I}     &       -0.064  &       0.221   \\
2477.87 &       \ion{Fe}{I}     &       -0.014  &       0.110   \\
2502.58 &       \ion{Si}{I}     &       -0.030  &       0.121   \\
2507.99 &       \ion{Mg}{I}     &       0.009   &       0.168   \\
2508.40 &       \ion{Mg}{I}     &       0.035   &       0.205   \\
2508.40 &       \ion{Mg}{I}     &       0.052   &       0.144   \\
\hline
\end{tabular}
\begin{tablenotes}
\item\textsuperscript{a} observed (1-F) at wavelength 
\item\textsuperscript{b} VALD3 stellar scaled to solar values
\end{tablenotes}
\end{table}

\begin{table}
\caption{Lines with cdepth>0.10 not present in the observed L- and M-band spectra.} 
\label{table:15}    
\begin{tabular}{c c c c}     
\hline\hline                       
Wavelength&species&Absorptance\tnote(\textsuperscript{a}&Synthetic\tnote(\textsuperscript{b} \\ 
(nm)            &ID                     &1-F             &   cdepth   \\
\hline
3473.12 &       \ion{Mg}{I}     &       -0.005  &       0.134   \\
3653.47 &       \ion{Mg}{I}     &       0.021   &       0.216   \\
3653.48 &       \ion{Mg}{I}     &       0.018   &       0.161   \\
3653.51 &       \ion{Mg}{I}     &       0.006   &       0.173   \\
3682.04 &       \ion{Fe}{I}     &       0.005   &       0.242   \\
3753.46 &       \ion{Mg}{I}     &       0.007   &       0.202   \\
3753.48 &       \ion{Mg}{I}     &       0.002   &       0.146   \\
3753.51 &       \ion{Mg}{I}     &       0.005   &       0.158   \\
3816.43 &       \ion{Mg}{I}     &       -0.001  &       0.112   \\
3817.51 &       \ion{Mg}{I}     &       0.002   &       0.176   \\
3885.37 &       \ion{Mg}{I}     &       -0.003  &       0.191   \\
3885.39 &       \ion{Mg}{I}     &       0.008   &       0.135   \\
3885.42 &       \ion{Mg}{I}     &       -0.002  &       0.147   \\
3886.01 &       \ion{Fe}{I}     &       0.000   &       0.157   \\
3891.67 &       \ion{Fe}{I}     &       0.007   &       0.138   \\
3895.44 &       \ion{Fe}{I}     &       -0.009  &       0.123   \\
3956.78 &       \ion{Mg}{I}     &       0.003   &       0.103   \\
3957.94 &       \ion{Mg}{I}     &       0.001   &       0.166   \\
4065.62 &       \ion{Mg}{I}     &       0.008   &       0.125   \\
4065.66 &       \ion{Mg}{I}     &       -0.004  &       0.138   \\
4093.97 &       \ion{Fe}{I}     &       0.000   &       0.161   \\
4114.68 &       \ion{Fe}{I}     &       0.003   &       0.190   \\
4120.60 &       \ion{Fe}{I}     &       -0.001  &       0.132   \\
4150.98 &       \ion{Mg}{I}     &       0.014   &       0.156   \\
4568.96 &       \ion{Si}{I}     &       -0.013  &       0.127   \\
4806.30 &        'CO'   &       -0.014  &       0.027   \\
4864.63 &       \ion{Mg}{I}     &       -0.030  &       0.107   \\
5013.08 &        'CO'   &       -0.009  &       0.027   \\
\hline
\end{tabular}
\begin{tablenotes}
\item\textsuperscript{a} observed (1-F) at wavelength 
\item\textsuperscript{b} VALD3 stellar scaled to solar values
\end{tablenotes}
\end{table}

\end{appendix}
\end{document}